\documentclass[%
 reprint,
 amsmath,amssymb,
 aps,
]{revtex4-2}
\usepackage[colorlinks=true, urlcolor=blue ]{hyperref}
\usepackage{graphicx}

\usepackage{dcolumn}
\usepackage{bm}

\usepackage{amsmath, amsthm, amssymb }
\usepackage{float,epsfig}
\usepackage{color}
\usepackage{multirow}
\usepackage{cleveref}
\usepackage{braket}
\usepackage{subfigure}

\usepackage{natbib}

\begin{document}

\newtheorem{theorem}{\bf Theorem}[section]
\newtheorem{proposition}[theorem]{\bf Proposition}
\newtheorem{definition}[theorem]{\bf Definition}
\newtheorem{corollary}[theorem]{\bf Corollary}
\newtheorem{example}[theorem]{\bf Example}
\newtheorem{exam}[theorem]{\bf Example}
\newtheorem{remark}[theorem]{\bf Remark}
\newtheorem{lemma}[theorem]{\bf Lemma}
\newtheorem{statement}[theorem]{\bf Statement}
\newcommand{\nrm}[1]{|\!|\!| {#1} |\!|\!|}

\newcommand{\calL}{{\mathcal L}}
\newcommand{\calX}{{\mathcal X}}
\newcommand{\calA}{{\mathcal A}}
\newcommand{\calB}{{\mathcal B}}
\newcommand{\calC}{{\mathcal C}}
\newcommand{\calK}{{\mathcal K}}
\newcommand{\C}{{\mathbb C}}
\newcommand{\R}{{\mathbb R}}
\newcommand{\U}{{\mathrm U}}
\renewcommand{\SS}{{\mathbb S}}
\newcommand{\LL}{{\mathbb L}}
\newcommand{\st}{{\star}}
\def\kernel{\mathop{\rm kernel}\nolimits}
\def\sigan{\mathop{\rm span}\nolimits}

\newcommand{\klasse}{{\boldsymbol \Delta}}

\newcommand{\ba}{\begin{array}}
\newcommand{\ea}{\end{array}}
\newcommand{\von}{\vskip 1ex}
\newcommand{\vone}{\vskip 2ex}
\newcommand{\vtwo}{\vskip 4ex}
\newcommand{\dm}[1]{ {\displaystyle{#1} } }

\newcommand{\be}{\begin{equation}}
\newcommand{\ee}{\end{equation}}
\newcommand{\beano}{\begin{eqnarray*}}
\newcommand{\eeano}{\end{eqnarray*}}
\newcommand{\inp}[2]{\langle {#1} ,\,{#2} \rangle}
\def\bmatrix#1{\left[ \begin{matrix} #1 \end{matrix} \right]}
\def \noin{\noindent}
\newcommand{\evenindex}{\Pi_e}

\newcommand{\tb}[1]{\textcolor{blue}{ #1}}
\newcommand{\tm}[1]{\textcolor{magenta}{ #1}}
\newcommand{\tre}[1]{\textcolor{red}{ #1}}



\def \K{{\mathbf k}}
\def \N{{\mathbb N}}
\def \R{{\mathbb R}}
\def \F{{\mathbb F}}
\def \C{{\mathbb C}}
\def \Q{{\mathbb Q}}
\def \Z{{\mathbb Z}}
\def \I{{\mathbb I}}
\def \D{{\mathcal D}}
\def \H{{\mathcal H}}
\def \P{{\mathcal P}}
\def \M{{\mathcal M}}
\def \B{{\mathcal B}}
\def \O{{\mathcal O}}
\def \calG{{\mathcal G}}
\def \PO{{\mathcal {PO}}}
\def \X{{\mathcal X}}
\def \Y{{\mathcal Y}}
\def \calW{{\mathcal W}}
\def \pf{{\bf Proof: }}
\def \lam{{\lambda}}
\def\lc{\left\lceil}   
\def\rc{\right\rceil}
\def \N{{\mathbb N}}
\def \Ls{{\Lambda}_{m-1}}
\def \Gb{\mathrm{G}}
\def \Hb{\mathrm{H}}
\def \Delta{\triangle}
\def \Rar{\Rightarrow}
\def \p{{\mathsf{p}(\lam; v)}}

\def \D{{\mathbb D}}

\def \tr{\mathrm{Tr}}
\def \cond{\mathrm{cond}}
\def \lam{\lambda}
\def \sig{\sigma}
\def \sign{\mathrm{sign}}

\def \ep{\epsilon}
\def \diag{\mathrm{diag}}
\def \rev{\mathrm{rev}}
\def \vec{\mathrm{vec}}

\def \ham{\mathsf{Ham}}
\def \herm{\mathsf{Herm}}
\def \sym{\mathsf{sym}}
\def \odd{\mathsf{sym}}
\def \en{\mathrm{even}}
\def \rank{\mathrm{rank}}
\def \pf{{\bf Proof: }}
\def \dist{\mathrm{dist}}
\def \rar{\rightarrow}

\def \bS{{\bf S}}
\def \cA{{\cal A}}
\def \E{{\mathcal E}}
\def \X{{\mathcal X}}
\def \F{{\mathcal F}}
\def \cH{\mathcal{H}}
\def \cJ{\mathcal{J}}
\def \tr{\mathrm{Tr}}
\def \range{\mathrm{Range}}
\def \adj{\star}

\preprint{APS/123-QED}


\title{Infinite reduction in absorbing time in quantum walks over classical ones}
\author{Shuva Mondal$^1$}
\email{shuvamondal@hri.res.in}
\author{Amrita Mandal$^{1,2}$}
\email{mandalamrita55@gmail.com, amrita.math@faculty.nita.ac.in}
\author{Ujjwal Sen$^{1}$}
\email{ujjwal@hri.res.in}

\affiliation{$^1$Harish-Chandra Research Institute, A CI of Homi Bhabha National Institute, Chhatnag Road, Jhunsi,  Prayagraj 211019, India}

 \affiliation{$^2$National Institute of Technology Agartala, Agartala 799046, India}


\begin{abstract}
We study the absorption time and spreading rate of the discrete-time quantum walk propagating on a line in the presence or absence of an absorber. We analytically establish that in the presence of an absorber, the average absorption time of the quantum walker is finite, contrary to the behavior of a classical random walker, indicating an infinite resource reduction on moving over to a quantum version of a walker. Furthermore, numerical simulations indicate a reversal of this behavior due to the insertion of disorder in the walker's step lengths. 
Additionally, we demonstrate that in the presence of an absorber, there is a speed-up in the spreading rate, and that a disordered quantum walk that is sub-ballistic regains the ballistic spreading of a clean quantum walk.



\end{abstract}

\maketitle



\section{Introduction}\label{sec:intro}
 Over the past few decades, there have been several developments in the area of quantum information and computation, for example, quantum dense coding, quantum algorithms, etc. \cite{kendon2006random, montanaro2016quantum}, and quantum walk is one of them that provides a universal quantum computational framework 
 \cite{childs_universal_2009,childs2013universal}.  
    The coined quantum random walk (QRW) model comprises a walker and a coin that controls the movement of the walker~\cite{first_QRW,nayak2000quantum,kempe2003quantum,ambainis2003quantum}. QRWs differ from classical random walks (CRWs) through different features and characteristics such as superposition, interference, and reversible unitary time evolution \cite{norris1998markov,spitzer2001principles,childs2002example,rudnick2004elements}. As a result of these distinctive characteristics, quantum walks show ballistic spreading, where the walker's position expands with time or the number of steps \cite{konno2002quantum,kendon2003decoherence}. In contrast to this, the spreading rate of classical random walks is diffusive in nature and is proportional to the square root of time  \cite{ambainis_bach_nayak_viswanath_watrous_2001}. The consequences of the ballistic nature of QRWs show diverse applications in the areas including quantum search algorithms \cite{shenvi2003quantum,portugal2013quantum,kadian2021quantum, apers2022quadratic},  triangle finding problem \cite{magniez2007quantum}, evaluating NAND trees \cite{jensen2019molecular,wang2020integrated}, and a platform for studying topological phases \cite{kitagawa2010exploring,dadras2018realization,xie2020topological} with applications involving ultra-cold rubidium atoms. In addition, the classical counterpart of QRW is a well-studied topic in various fields, such as the modeling of Brownian motion \cite{kac1947random,romanelli2005decoherence}, estimating the volume of a convex body using Markov chain models \cite{chakrabarti2023quantum}, modelling boson hopping~\cite{PhysRevA.96.043629,anisur2025directional}, and the PageRank algorithm 
 \cite{ PhysRevLett.108.230506,chawla2020discrete}.
 
 QRWs have been extensively studied for both discrete \cite{meyer1996from,meyer1996absence} and continuous time steps \cite{strauch2006connecting,mulken2011continuous}. In this paper, we focus on the former one.  In discrete-time quantum walks (DTQWs), at each time step, the walker moves through discrete lattice points \cite{lovett2010universal, childs2010relationship, PhysRevA.83.062315}. The evolution of the walker of a DTQW can be hindered by the presence of different unavoidable environmental effects, such as disorder or noise \cite{PhysRevA.86.022335,PhysRevA.89.042307}. The term disorder in a quantum system refers to the introduction of randomness or imperfections into the system, potentially having a substantial effect on the behavior and properties of the quantum particle. Quantum walks with disorder have been studied by considering disorders in the displacement operations \cite{article,article1,sreetama_di, PhysRevResearch.2.023002, PhysRevE.102.012104, PhysRevA.106.042408} or in the coin operations \cite{salimi2012asymptotic,vieira2013dynamically,rohde2013quantum,vieira2014entangling,di2016discrete,di2016discrete,montero2016classical,wang2018dynamic,orthey2019weak,singh2019accelerated,buarque2019aperiodic,pires2021negative,priya_2021,naves2023quantum}. The presence of disorder exhibits significant influence on the spreading behavior of the walker, that generally inhibits the ballistic spreading and results in partial or complete localization \cite{linden2009inhomogeneous}. 
 
 Disorder can be added into the system by introducing random variations in the potential landscape or the hopping amplitudes, or it may appear due to the unavoidable interaction between the system and the environment. In the context of QRWs, the disorder can be categorized into two distinct forms: static disorder and dynamic disorder \cite{chatterjee1994effective,yin2008quantum,nosrati2021readout}. Static disorder involves a position-dependent random variable that represents a fixed, random arrangement of impurities in a lattice, which remains constant over time. The randomized parameters in static disorder suppress quantum evolution, which can lead to phenomena like Anderson localization \cite{mendes2021localization,mandal2023localization,duda2023quantum}.  The dynamic disorder is characterized by a position- and time-dependent random variable, which involves fluctuations or vibrations in the lattice over time, resulting in decoherence. This induces a quantum-to-classical transition \cite{yin2008quantum,vieira2014entangling,ampadu2014some,nosrati2021readout, PhysRevA.109.022224}. The presence of an absorber can have a substantial influence on the dynamics of the QRWs as well. If the quantum particle is detected at the absorber's site, it gets absorbed, thereby terminating the walk \cite{yamasaki2003analysis, bach2004one, PhysRevA.84.032319}.
     Absorber introduces a non-unitary effect into the evolution of quantum walk, which results into interesting phenomena like as absorption probability and exists probability\cite{kuklinski2018absorption, PhysRevA.101.032309}.

     The rest of this paper is organized as follows. Section~\ref{sec:DTQW on a line} introduces the theoretical framework for quantum walks and the spreading rate of walks. In Section~\ref{sec:absorbing time} the effect of an absorber in DTQW is described. There we also presented our 1st result about finite average absorbing time for DTQW in contrast to classical random walk in the presence of an absorber. There we also presented our numerical findings on how things get reversed if there is disorder in the lengths of steps of the walker. Section~\ref{sec:spreading rate} presents how the presence of an absorber affects the spreading rate of a walker even when the steps of the walker are disordered. Finally, Section~\ref{sec:conclusion} concludes the paper and outlines future research directions.

\section{DTQW on a line}\label{sec:DTQW on a line}
In this section, we briefly discuss the discrete-time quantum walks (DTQWs) on a line. In DTQW, the system is described by a state vector defined in the Hilbert space $\mathcal{H}=\mathcal{H}_{p}\otimes \mathcal{H}_{c},$ where the Hilbert spaces $\mathcal{H}_{c}$ and $\mathcal{H}_{p}$ denote the coin and the position spaces, respectively. 
The Hilbert space $\mathcal{H}_{c}=\mathrm{span}\left\{ \ket{L}, \ket{R}\right\}$ describes the two-dimensional coin space where $\{\ket{L},\ket{R}\}$ forms the canonical ordered basis for $\mathbb{C}^2$ and, $L$ and $R$ stand for left and right directions of the walker. The infinite-dimensional position space $\mathcal{H}_p$ is defined as $\mathcal{H}_{p}=\mathrm{span}\left\{\ket{m}|m\in \mathbb{Z}\right\},$ where $\ket{m}$ indicates the position vector of the walker on the 1D lattice $\mathbb{Z}$. Hence, the total Hilbert space is $\mathcal{H}=\mathrm{span}\left\{\ket{m,l}|m\in \mathbb{Z},l\in \{L,R\}\right\}.$ The DTQW propagates through the lattice by performing two operations on the walker's state at each time step, the coin tossing operation $C$ followed by the shift operation ${S}.$ Thus, the evolution operator for the walk takes the form 
\begin{equation} \label{eq: def_of_basic_walk}
{W}={S}(I\otimes {C}),
\end{equation} where $I$ is the identity operator acting on $\mathcal{H}_p.$ Hence, after $t$ time steps, the state of the walker becomes $\ket{\psi_t}={W}^t \ket{\psi_0},$ where $\ket{\psi_0}$ is the initial state of the walker.

The action of the coin operator {\scriptsize ${C}=\bmatrix{a&b\\c&d}$} is described as,
${C}\ket{L}=a\ket{L}+b\ket{R},{C}\ket{R}=c\ket{L}+d\ket{R},$ where $\{a,b,c,d\} \in \mathbb{C}$ are chosen accordingly depending on the choice of the coin operator. For example, in the case of the Hadamard walk, that is described by the single qubit Hadamard gate $H$ as the coin operator,
\begin{equation} \label{eq:Hadamard params}
a=-b=-c=-d=-\frac{1}{\sqrt{2}}.
\end{equation} The shift operator $S$ is defined as \begin{equation}\label{eq: def_shift_opr}
     {S}=\sum_{n=-\infty}^{\infty}\ket{n-1,L}\bra{n,L}+\ket{n+1,R}\bra{n,R},
 \end{equation} and its action on the basis states is given by ${S}\ket{n, L}=\ket{n-1, L},
{S}\ket{n,R}=\ket{n+1,L}.$ 
Finally, from Eq.~(\ref{eq: def_of_basic_walk}) we get
\begin{eqnarray}\label{eq: action_of_basic_walk}
{W}\ket{n,L}&=&a\ket{n-1,L}+b\ket{n+1,R} \nonumber\\
{W}\ket{n,R}&=&c\ket{n-1,L}+d\ket{n+1,R}.
\end{eqnarray} 

Let the initial state of the walk be expanded as
$$\ket{\psi_0}=\sum_{n}\psi^L_0(n)\ket{n,L}+\psi^R_0(n)\ket{n,R},$$ where $\psi^L_0(n)$ and  $\psi^R_0(n)$ are the probability amplitudes of the coin states $\ket{L}$ and $\ket{R},$ respectively, at position $n$ and at time $t=0.$
Then, by applying $W$ on $\ket{\psi_0}$ for $t$ times, we get
\beano \ket{\psi _{t}}&=&{W}^{t}\left(\sum_{n}\psi^L_0(n)\ket{n,L}+\psi^R_0(n)\ket{n,R}\right)\\
&=&\sum_{n}\psi^L_t(n)\ket{n,L}+\psi^R_t(n)\ket{n,R}.\eeano
Hence, the probability of the walker to be at the position $n$ after time $t$ is given by
\beano p(n,t)&=&|\langle{n,L}|{\psi _{t}}\rangle|^{2}+|\langle{n,R}|{\psi _{t}}\rangle|^{2}\\
&=&|\psi^L_t(n)|^2+|\psi^R_t(n)|^2.\eeano
The standard deviation $\sigma,$ that quantifies the spreading rate of the walk, can be expressed in terms of the position probability distribution $p(n,t)$ as follows.  
$$\sigma=\sum_{n}n^{2}p(n,t)-\left(\sum_{n}n p(n,t)\right)^{2}.$$ 
Now the dependence of $\sigma$ on time $t$ can be scaled by the exponent $\alpha$ as
\begin{equation} \label{eq: def_of_exponent}
    \sigma \propto t^{\alpha}.
\end{equation}
From Eq.~(\ref{eq: def_of_exponent}), it is evident that $\alpha$ describes how the nature of the walker spreading changes with time. For example~\cite{kempe2003quantum}, $\alpha$ takes value $1$ in clean DTQWs.

\section{Infinite absorbing time in classical walk vs finite duration in the quantum case
} \label{sec:absorbing time}
In this section, we review the effects on the evolution of the clean CRW due to the presence of the absorber and analyze the same for clean DTQW. 
Physically, the walker stops its propagation while crossing the absorber that is present on the line of the walker's propagation, and the probability of moving to the subsequent position beyond the absorber gets absorbed. 
So, the walker can not move to the other side of the absorber. The probability to be at the other side of the absorber and the probability to be at the position of the absorber are lost.

Let $p_{t}$ denotes the probability of being absorbed for the first time after $t$ tosses at $m_{1},$ the position of the absorber. Then, the total probability of absorption after considering all the time steps is 
\begin{equation}\label{eq: total absorption probability}
    P=\sum_{t=1}^{\infty} p_{t}.
\end{equation}
Let us denote $t_a$ as the average absorbing time, and it is defined as
\begin{equation}\label{eq: def_of_avg_absorbing_time}
    t_{a}=\frac{\sum_{t=1}^{\infty}t p_{t}}{\sum_{t=1}^{\infty} p_{t}}.
\end{equation}
Note that the total probability of absorption $P$ needs not be $1$, and hence, we have to divide with $P$ to define $t_a$ in a meaningful way. In the following, we discuss the quantities (\ref{eq: total absorption probability}) and (\ref{eq: def_of_avg_absorbing_time}) for both classical and quantum cases in detail.

\subsection{Classical walk} \label{subsec: class}
In CRW, if the distance of the absorber from the starting position is $|m_{1}|$, then $p_{t}$ is nonzero only if $t\geq m_{1},$ and $\frac{t+m_1}{2}$ is an integer. From \cite{Chandrashekhar_classical}, we get
\beano
p_{t}=\frac{\lvert m_{1}\lvert}{t\times 2^{t}}\frac{t!}{(\frac{t+m_{1}}{2})!(\frac{t-m_{1}}{2})!}.
\eeano
Further, it can be shown that $P$ as described in Eq.~(\ref{eq: total absorption probability}), is equal to $1$ for CRW (see Appendix~\ref{apndx B}). So that, for a classical random walk, the expression of the average time of absorption becomes
\begin{equation} \label{eq: average_absorbing_time_in_classical}
    t_{a}=\sum_{t}t p_{t}=\sum_{t}\frac{\lvert m_{1}\lvert}{2^{t}}\frac{t!}{(\frac{t+m_{1}}{2})!(\frac{t-m_{1}}{2})!},
\end{equation}
where if $m_{1}$ is even (odd), the sum over $t$ ($\geq m_{1}$) runs over the even (odd) numbers. We show that $t_{a}$ is not finite for a clean CRW by proving the divergence of the infinite series sum (see Appendix~\ref{apndx C} for the detail).

\subsection{Quantum walk}

In DTQW, the absorption probabilities ($p_{t}$) can be derived from amplitudes $a_{t}$ of getting absorbed after time $t$ by the relation
$p_{t}=|a_{t}|^{2}.$
Now these $a_{t}$ can be calculated by defining the generating function $G(z)$ described as
\begin{equation}\label{eq: calculation of amplitude}
    G(z)= \sum_{t>0} a_{t} z^{t},
\end{equation}
where $z$ is a complex variable.
This $G(z)$ depends on the walker's initial state, the coin operator, and the position of the absorber. For example,  consider DTQW  with $H$ as the coin operator, and the absorber is at the position $m_{1}.$ Then, if $\ket{0, R}$ is the initial vector, we get $G(z)=(g(z))^{m_{1}},$ otherwise, if $\ket{0, L}$ is the initial state, $G(z)=f(z)(g(z))^{m_{1}-1}$,\cite{bach2002onedimensional} where
\begin{align*}
    f(z)=\frac{1+z^{2}-\sqrt{1+z^{4}}}{\sqrt{2}z},\\
g(z)=\frac{1-z^{2}-\sqrt{1+z^{4}}}{\sqrt{2}z}.
\end{align*}
Here, we have assumed $m_{1}>0$, but $m_{1}<0$, then $f(z)$ and $g(z)$ exchange their aforesaid expressions. From now onwards, unless not stated explicitly, we use $H$ as coin, $\ket{0, L}$ as the initial state, and $m_1=2$ as the position of the absorber, so that when the walker gets absorbed its state is described as $\ket{2, R}$.  Under such circumstances, $G(z)$ in Eq.~(\ref{eq: calculation of amplitude}) simplifies to 
\begin{equation} \label{eq: f for hadamard not expanded}
    G(z)=\frac{1-\sqrt{1+z^{4}}}{z^{2}}.
\end{equation}
Now, expanding Eq.~(\ref{eq: f for hadamard not expanded}) in powers of $z$ we get,
\begin{equation}\label{f for hadamard expanded}
G(z)=\sum_{m=1}^{\infty}(-1)^{m}\frac{[2(m-1)]!}{2^{2m-1}(m-1)!m!}z^{4m-2}.
\end{equation} 
It is clear from the above power series that we get nonzero amplitude only when the corresponding toss number $t$ equals to $4m-2,$ and hence, we write the expression of the amplitudes by introducing the Kronecker delta function as 
\beano
a_{t}=\sum_{m \in \mathbb{N}}(-1)^{m}\frac{[2(m-1)]!}{2^{2m-1}(m-1)!m!} \delta_{t,(4m-2).}
\eeano
So that the absorption probabilities become
\begin{equation} \label{eq: quantm_absorption_probabilities}
    p_{t}=\sum_{m\in \mathbb{N}}\left(\frac{[2(m-1)]!}{2^{2m-1}(m-1)!m!}\right)^{2} \delta_{t,(4m-2)}.
\end{equation}
Now, using (\ref{eq: total absorption probability}) the total absorption probability $P$ is written as
\beano
P=\sum_{m=1}^{\infty}\left(\frac{(2m-2)!}{2^{2m-1}(m-1)!m!}\right)^{2}.
\eeano
The series sum in the right-hand side of the above expression of $P$ converges and yields a total absorption probability value $P=\frac{4}{\pi}-1$. From (\ref{eq: def_of_avg_absorbing_time}), the average absorbing time $t_a$ becomes,
\begin{equation} \label{eq: quantum_avg_time}
    t_{a}= \frac{\sum_{m=1}^{\infty}(4m-2)(\frac{(2m-2)!}{2^{2m-1}(m-1)!m!})^{2}}{\sum_{m=1}^{\infty}(\frac{(2m-2)!}{2^{2m-1}(m-1)!m!})^{2}}.
\end{equation}
The infinite series in the numerator of Eq.~(\ref{eq: quantum_avg_time}) converges (See Appendix~\ref{apndx C} for details). Further, it can be derived that this series sum in the numerator takes the value $\frac{\pi -2}{\frac{\pi}{2}}.$ So that the average absorption time can be approximated as 
$t_{a}=\frac{\pi -2}{\frac{\pi}{2}(\frac{4}{\pi}-1)} \approx 2.66.$

It is clear that as the position of the absorber shifts, the total absorbing probability $P$ and the corresponding average absorbing time $t_a$ change as well. We collect all these data in Table~\ref{table:total absorbing probability variation}, considering the absorber's position at the lattice points $1,\ldots,10$ on the line. It is evident from the table that the total absorbing probability gradually decreases and the average absorbing time increases as the distance between the starting point and the position of the absorber increases.
In all such cases, we see that there is a linear relation between the average time of getting absorbed and the distance of the absorber from the starting position of the quantum walker. However, a reverse relation is observed for the total probability of getting absorbed with the absorber position. 
\begin{table}[ht]
\centering
\begin{tabular}{c c c}
\hline
$m_{1}$ & P & $t_{a}$ \\ [0.7ex] 
\hline \hline
$1$ & $0.64$ & $1.57$ \\
$2$ & $0.27$ & $2.66$ \\
$3$ & $0.18$ & $4.87$ \\
$4$ & $0.16$ & $7.48$ \\
$5$ & $0.15$ & $10.07$ \\
$6$ & $0.14$ & $12.91$ \\
$7$ & $0.14$ & $15.50$\\
$8$ & $0.14$ & $18.07$\\
$9$ & $0.14$ & $20.61$\\
$10$ & $0.14$ & $23.39$\\
\hline
\end{tabular}
\caption{Here $m_{1}, P$ and $t_{a}$ denote the absorber's position, the total probability of absorption, and the average time of getting absorbed at $m_1,$ respectively. These values are approximated up to two decimal places. It is to be noted that $t_a$ increases linearly with the distance of the absorber from the starting position of the quantum walker. Besides, an inverse relation is noticed between $m_1$ and $P.$
}
\label{table:total absorbing probability variation}
\end{table} 

An interesting quantum mechanical impact on the average absorbing time of a walker in the presence of an absorber is that the classical walker cannot get absorbed after finite time steps; in contrast, the quantum walker does the same in finite time whenever the absorber is placed at a finite distance from the starting position of the walker. We notice that the absorption probabilities ($p_t$) carry the same values and hence lead to almost similar results while using the Kempe coin \cite{kempe2003quantum}, instead of the Hadamard one.

\vspace{-2cm}
\subsection{Effects in the presence of disorders}
In this section, we discuss the effects on the average absorbing time by introducing disorder in the walker's steps, both in classical and quantum scenarios. First, for a single realization of the disorder insertion, we calculate the weighted average time quantity 
\begin{equation}\label{eq: def_of_avg_absorbing_time a}
    t_{a} ^{(n)}=\frac{\sum_{t=1}^{n}t p_{t}}{\sum_{t=1}^{n} p_{t}},
\end{equation} for a fixed $n.$ Then, we consider Eq.~(\ref{eq: def_of_avg_absorbing_time a}) for different values of $n,$ and for several such disorder realizations, we take the average of the quantity defined in Eq.~(\ref{eq: def_of_avg_absorbing_time a}) to get the disorder-averaged absorbing time $\langle t_{a}^{(n)}\rangle$ of the disordered walks. Here, the symbol \(\langle \cdot \rangle\) stands for an average of the argument over the corresponding disorder realizations.

\begin{figure}[H]
    \includegraphics[width=3.4 in, height=6cm]{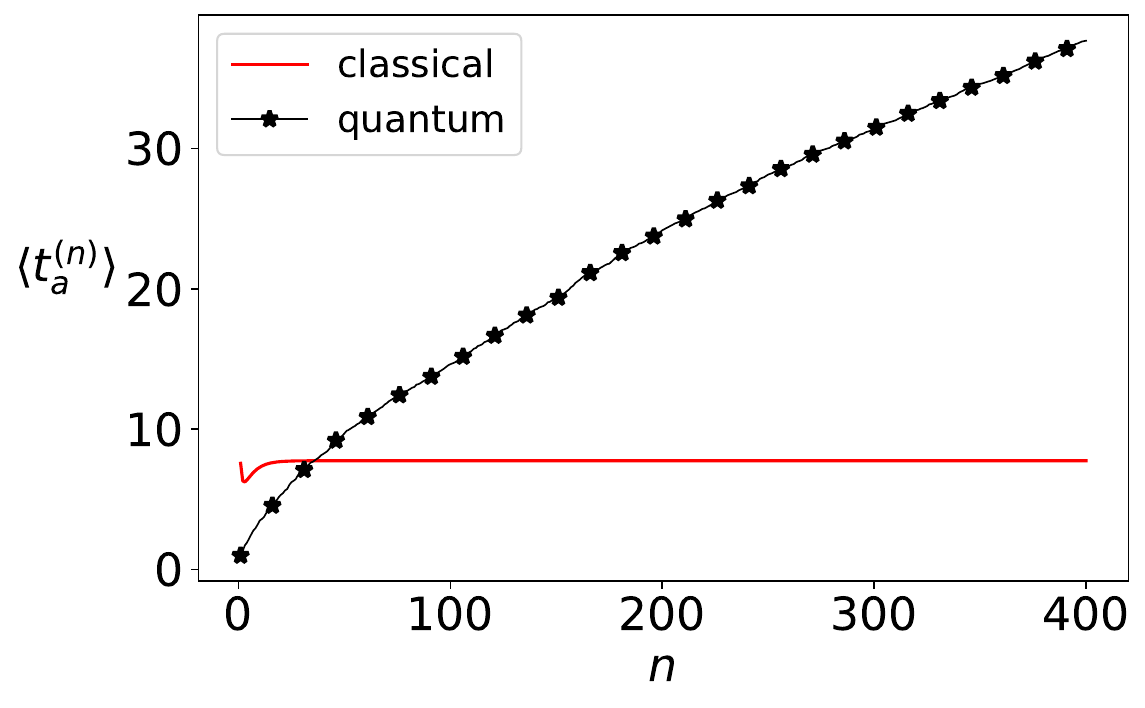}
    \caption{A comparison of the disorder-averaged absorbing time of CRW and DTQW, both with the presence of an absorber at the position $2$. The Poisson distribution with unit mean is the underlying disorder. The red line, without any marker, corresponds to the disordered-CRW. This plot shows a sudden fall at the beginning
    and no significant change is observed
    in the $\langle t_{a} ^{(n)}\rangle$ values after few initial steps. As $n$ increases, $\langle t_{a} ^{(n)}\rangle$ attends a constant value of $7.75$ (approximately), showing the convergence nature of  $(\langle t_{a} ^{(n)}\rangle)_n$ for the classical walk. Here, we have considered numerical simulations up to $n=400$ steps. The black, asterisk-marked plot shows the behavior of the disorder-averaged absorbing time of disordered-DTQW defined by the Hadamard coin. This plot exhibits the monotonically increasing nature of $(\langle t_{a} ^{(n)}\rangle)_n,$ within the considered range of $n$ values up to $400$ steps ($\langle t_{a} ^{(n)}\rangle\approx 38$ at $n=400$). This signifies that $(\langle t_{a} ^{(n)}\rangle)_n$ diverges for the DTQW we considered. 
    Here, the averaged time quantity is derived over $40$ different disorder realization runs.
    } \label{fig:classical-disordered-absorbing-time}
\end{figure}

We plot the $n$ values along the horizontal axis and corresponding disorder-averaged time quantity $\langle t_{a} ^{(n)}\rangle$ along the vertical axis for disordered-CRW and disordered-DTQW in Fig.~\ref{fig:classical-disordered-absorbing-time}. Here, we have considered discrete random outputs that follow the Poisson distribution with unit mean as the disorder insertion. The red line without any marking, representing disordered-CRW in Fig.~\ref{fig:classical-disordered-absorbing-time}, clearly illustrates that there is no significant change in the values of $\langle t_{a} ^{(n)}\rangle$ after a certain finite steps. This indicates that for the classical walk, the sequence of disorder-averaged absorbing time $(\langle t_{a} ^{(n)}\rangle)_n$ converges to a finite value with respect to $n.$ On the other hand, the black asterisk marked line in Fig.~\ref{fig:classical-disordered-absorbing-time} depicts an increase in $n$ value resulting in a gradual increment of $\langle t_{a} ^{(n)}\rangle$, when the disordered steps are chosen from the Poisson distribution with the specified parameters. This clearly shows the divergent nature of the sequence $(\langle t_{a} ^{(n)}\rangle)_n$ with respect to the step size $n,$ for the quantum case, when the disordered steps are collected from the Poisson distribution with unit mean.

 We have also found similar behavior for $\langle t_{a} ^{(n)}\rangle$ in disordered-DTQW, when the disordered step lengths are chosen from the other four types of discrete probability distribution, i.e., binomial, negative binomial, geometric, and hypergeometric distributions with different parameter setup. In all such cases,  $\langle t_{a} ^{(n)}\rangle$ increases monotonically with $n$. This enlargement in $\langle t_{a} ^{(n)}\rangle$ demonstrates the divergent nature of the disorder-averaged absorbing time, analogous to the Poisson disorder-averaged absorbing time in disordered-DTQW.
Thus, considering all the discussions, we state that the behavior of the average time of absorption is altered in the presence of disorders for both classical and quantum walks.

\section{Absorber renders disorder irrelevant in spreading rate of quantum walk} \label{sec:spreading rate}
In this section, we study the impacts of an absorber on the spreading rate of DTQWs propagating in a 1D line. Additionally, we execute a comparative study of the spreading behavior of CRWs and DTQWs under such circumstances. In the present study, the spreading rate is quantified in terms of the exponent $\alpha$ as stated in Eq.~(\ref{eq: def_of_exponent}). 
It is well known that in a clean CRW, where no absorber is present along the path of the walker, the probabilities show a diffusive normal distribution with standard deviation 
$\sigma\propto t^\frac{1}{2},$ and therefore $\alpha=0.50$ \cite{Chandrashekhar_classical}.  At the same time, for a clean DTQW, where an absorber is not present, $\alpha$ is exactly equal to $1$ and the walk is called ballistic in nature \cite{portugal2013quantum}.
This quadratic speed up in the quantum regime exhibits the quantum advantages of random walks over the classical ones. 

We observe that insertion of several disorders into the step lengths results in no significant change in the spreading of the 1D CRWs, and $\alpha$ almost remains $0.50.$
On the other hand, it has been demonstrated in \cite{sreetama_di} that for a disordered DTQW, the exponent $\alpha$ reduces noticeably and lies between $\frac{1}{2}$ and $1$ Hence, the DTQW becomes sub-ballistic but super-diffusive in nature.
In particular, for the Poissonian disorder (see Appendix \ref{sec:glassy_disorder}) insertion, if we take an average of several runs of the unit mean Poisson-distributed disorders, $\alpha$ drops to $\alpha\approx0.70$ from $\alpha=1$. So, even though the spreading of a CRW remains unaffected by step-length disorders, certain changes in the propagation of quantum walks are observed.

\begin{figure}[h]
    \centering
    \includegraphics[width=\linewidth, height=8cm]{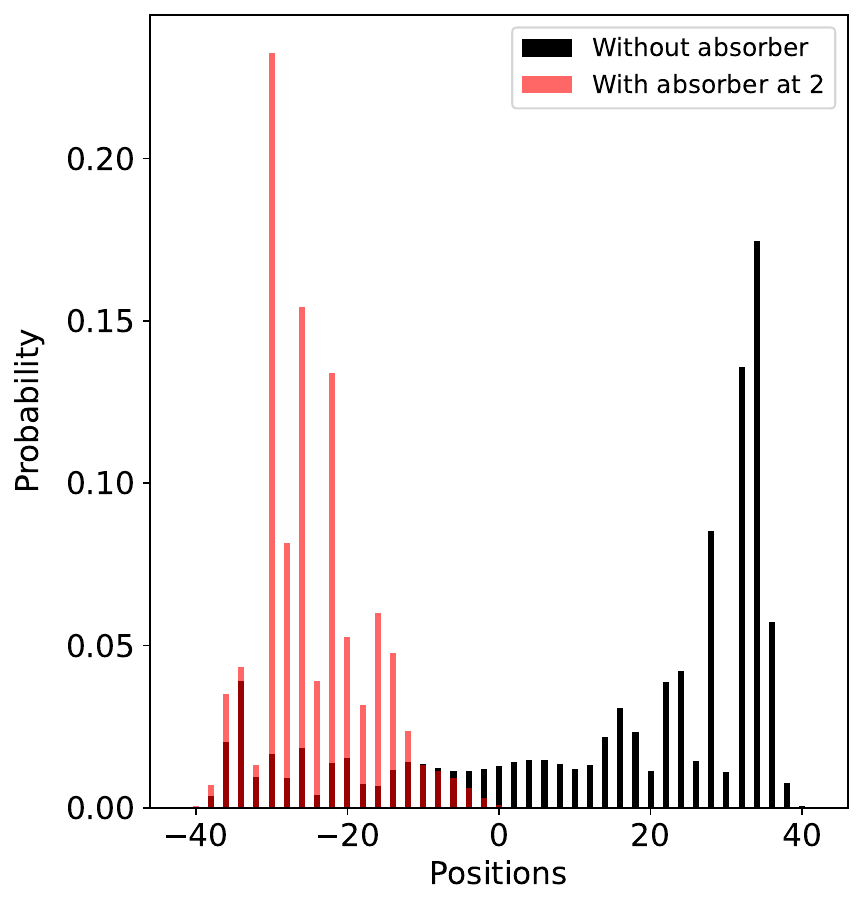}
    \caption{Probability distribution of 1D-DTQW. This position-probability plot describes the probability distribution nature of the 1D-DTQW after $50$ time steps, starts with the initial state $\ket{0}\otimes \ket{0}.$ Here, the black vertical lines quantify the probability of the walker being at the specified positions on the line, when there is no absorber on the line. Here some higher probability values are observed along the positive nodes (right-hand side) compared to the other side. On the other hand, the red colour represents the case when there is an absorber at position $2.$ As a result of the absorber present at node $2$, most of the non-zero probabilities are noticed along the negative nodes, with the highest peak around $-32.$ }
    \label{fig:combined1}
\end{figure}

Now, we turn to the situation where an absorber is placed along the line of dispersion of the random walk. Then, the probability of existence of the particle should be drained. In order to obtain accurate findings, we need to adjust the remaining probability following the probability leakage due to the absorber.
 In the classical scenario, to make the modification, the probability at each position is divided by the total remaining probability, i.e., the total probability that the particle is not absorbed yet. The presence of the absorber does not have an impact on the spreading rate, and $\alpha$ remains $0.50$. In the quantum regime, we renormalize the wave function over the locations where the absorption does not occur. In Fig. \ref{fig:combined1}, we draw and compare the probability distributions of clean Hadamard 1D-DTQWs against some nodal positions, when the absorber is absent and present at a vertex of the line graph after some certain time-steps. In both cases, the walkers start from the position $0$ with initial state $\ket{0}$, and snapshots for the probability distributions are taken after $20$ times of evolution. Clearly, due to the presence of the absorber, a certain difference between the two position-probability distribution graphs is noticed, indicating different statistical characteristics of the distributions.

\begin{figure}[]
    \includegraphics[width=9.5 cm, height=7cm]{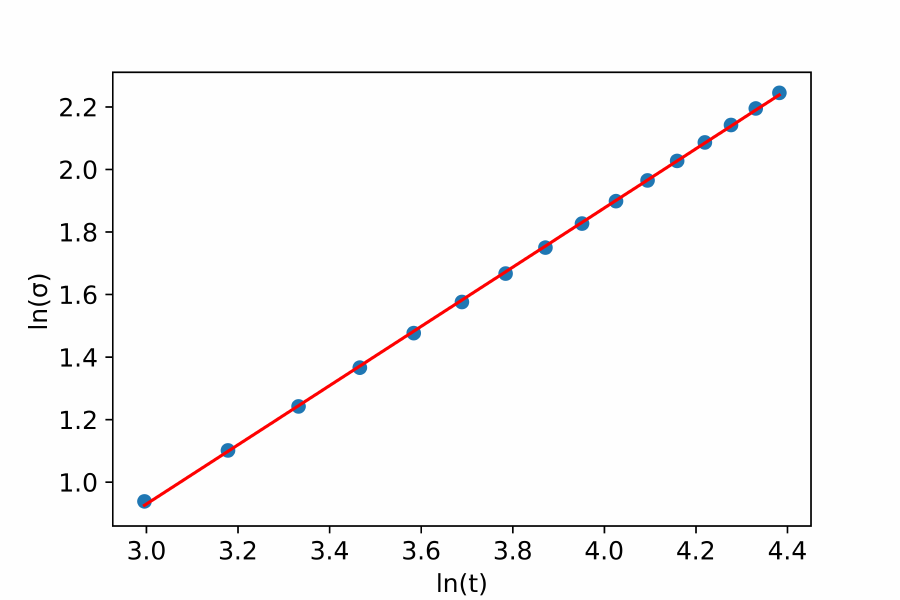}
    \caption{Plot of $\ln(\sigma)$ against $\ln(t)$.  The linear fitting (on the log-log plot) for the data points (blue colored dots) corresponding to the Hadamard 1D-DTQW with one absorber placed at $2$ is represented by red solid line.
    With $95\%$ confidence level, the slope of the fitted line falls within the confidence intervals
$0.96\pm 0.005$, with an average least square error of $0.002$ for the linear fittings. To ensure fewer errors in the fitting, we have chosen the $t$-range \(20 < t < 80\). 
    }
    \label{fig: with_abs_without_dis}
\end{figure}

Furthermore, we observe that for 1D-DTQW with the (single-qubit) Hadamard coin, the scaling exponent $\alpha$ decreases slightly from $1$ when an absorber is present. To determine this $\alpha$, we first compute and collect standard deviations ($\sigma$) for some values of time $t.$ Then, we create a graph of $\ln(\sigma)$ against $\ln(t)$ and fit it with a straight line with $95\%$ confidence level. The slope of this line now signifies the exponent $\alpha$, as defined in Eq.~(\ref{eq: def_of_exponent}). Figure \ref{fig: with_abs_without_dis} shows the plot $\ln(\sigma)$ versus $\ln(t)$ of 1D Hadamard walk for $t\leq 80,$ when the absorber is placed at the node $2$ on the line. This results in an exponent value of $0.96\pm 0.005$ with confidence level $95\%$, other than that of $\alpha=1$, which occurs for quantum walks in the absence of the absorber. This decreased $\alpha$ value is also observed for placing the absorber at other nodes as well, asserting the effectiveness of the absorber in the spread of the DTQWs.

\begin{figure}[h!]
    \centering
    \includegraphics[width=\linewidth, height=9cm]{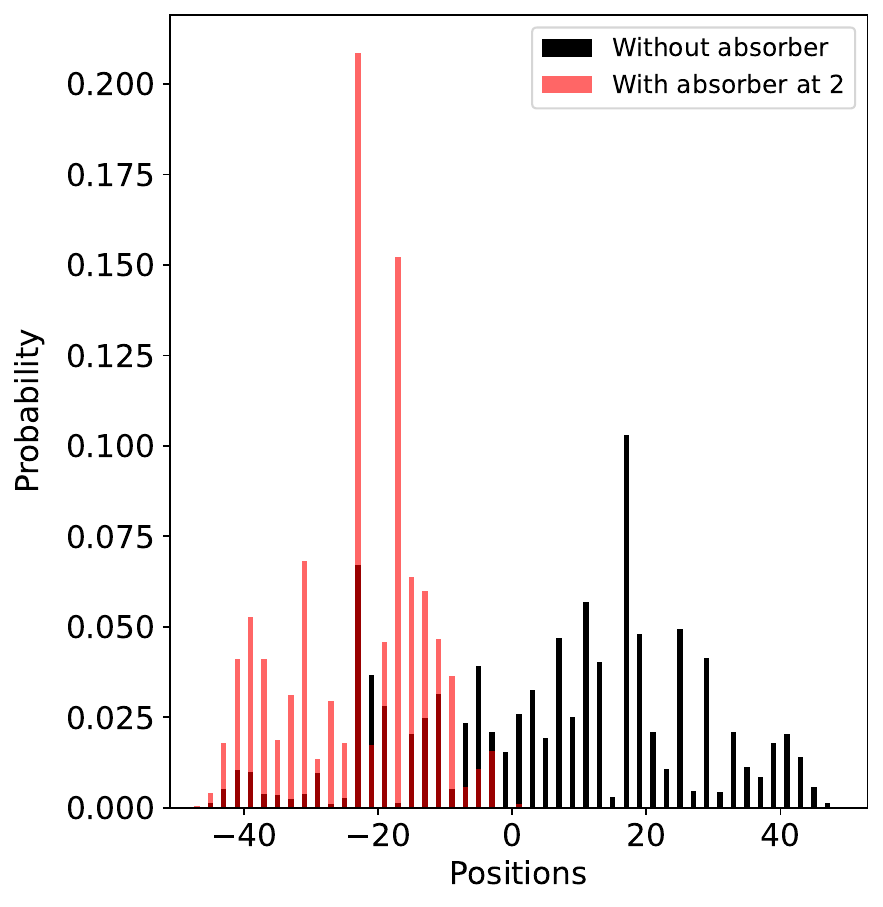}
    \caption{
    Diagram of probability distribution of 1D Poisson disordered-DTQW defined by Hadamard coin after $50$ evolutions, for a single disorder realization with mean value $1$, without and with the presence of the absorber.
    In both cases, the walker starts from $0$ node with initial coin state $\ket{0}$, and the outcomes of the Poisson distribution of unit mean in a single run are taken as the disorder realizations.
    The black color lines show probability distributions of disordered-DTQW over different locations, without the presence of any absorber. The red bars represent the position-probability distribution of disordered-DTQW, where an absorber is kept at node $2.$ In contrast to the no absorber case, non-zero probabilities are seen to be more localized near the starting position $0$ when the absorber is placed.}
    \label{fig:combined2}
\end{figure}

Finally, we undertake the condition where CRWs or 1D-DTQWs are treated in the presence of glassy disorders in the walkers' step lengths, and absorbers at a node on the walker's line of propagation. Under these circumstances, we study the walk statistics and provide a comparative view with all the results discussed earlier. Fig. \ref{fig:combined2}
represents a comparative view of the position-probability distributions after $20$ time steps of disordered 1D-DTQW defined by the single qubit Hadamard coin, without and with the presence of an absorber on the line graph. 
In both cases, the walkers start from the position $0$ and the initial coin state is $\ket{0}.$ 
Now, to predict the spreading rate of the disordered walk, we consider the walk in different runs, as stated in Section \ref{sec:absorbing time}. 
In each run, the walk is treated by taking steps of random lengths drawn from a set of integers that follow a given probability distribution with some fixed parametric values. We call this process a single disorder realization of the walk. 
Then, for several such disorder realizations, we numerically evaluate $\sigma$, the measure of spreading of the walk.  Then, we take the average of the $\sigma$s over all the disorder realizations to get $\langle\sigma\rangle$. 
Now, to find the spreading rate, we first plot $\ln(\langle\sigma\rangle)$ along the vertical against $\ln(t)$ along the horizontal and fit a straight line through the points with a minimum possible error of fitting, using the least squares method. The slope of the fitted straight line signifies the disorder-averaged spreading rate.


\begin{figure}[h!]
    \includegraphics[width=8.9cm]{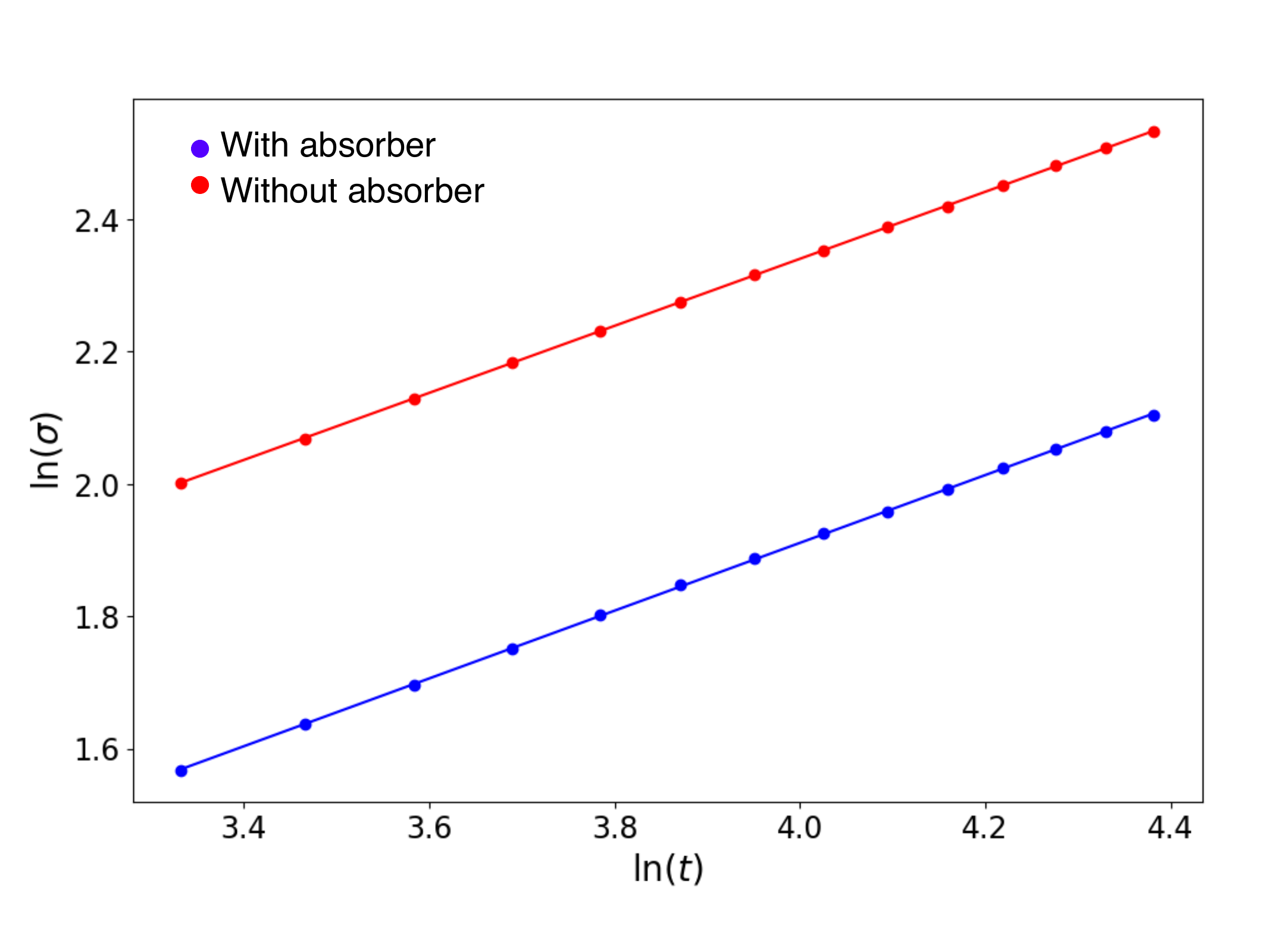}
    \caption{(Color online.) $\ln(\sigma)$ against $\ln(t)$ for disordered (Poisson distribution)  CRWs with and without absorber scenarios. Here, $t$ ranges between $20$ and $80.$ The red and blue fitted lines represent the cases without and with an absorber, respectively. Red line equation is $ \ln(\sigma) = 0.51 \ln(t) + 0.31,$ and the blue line is $ \ln(\sigma) = 0.51 \ln(t) -0.14$. The slopes of both lines are the same. } \label{fig: classical spread}
\end{figure}

\begin{figure}[]
    \includegraphics[width=8.9cm]{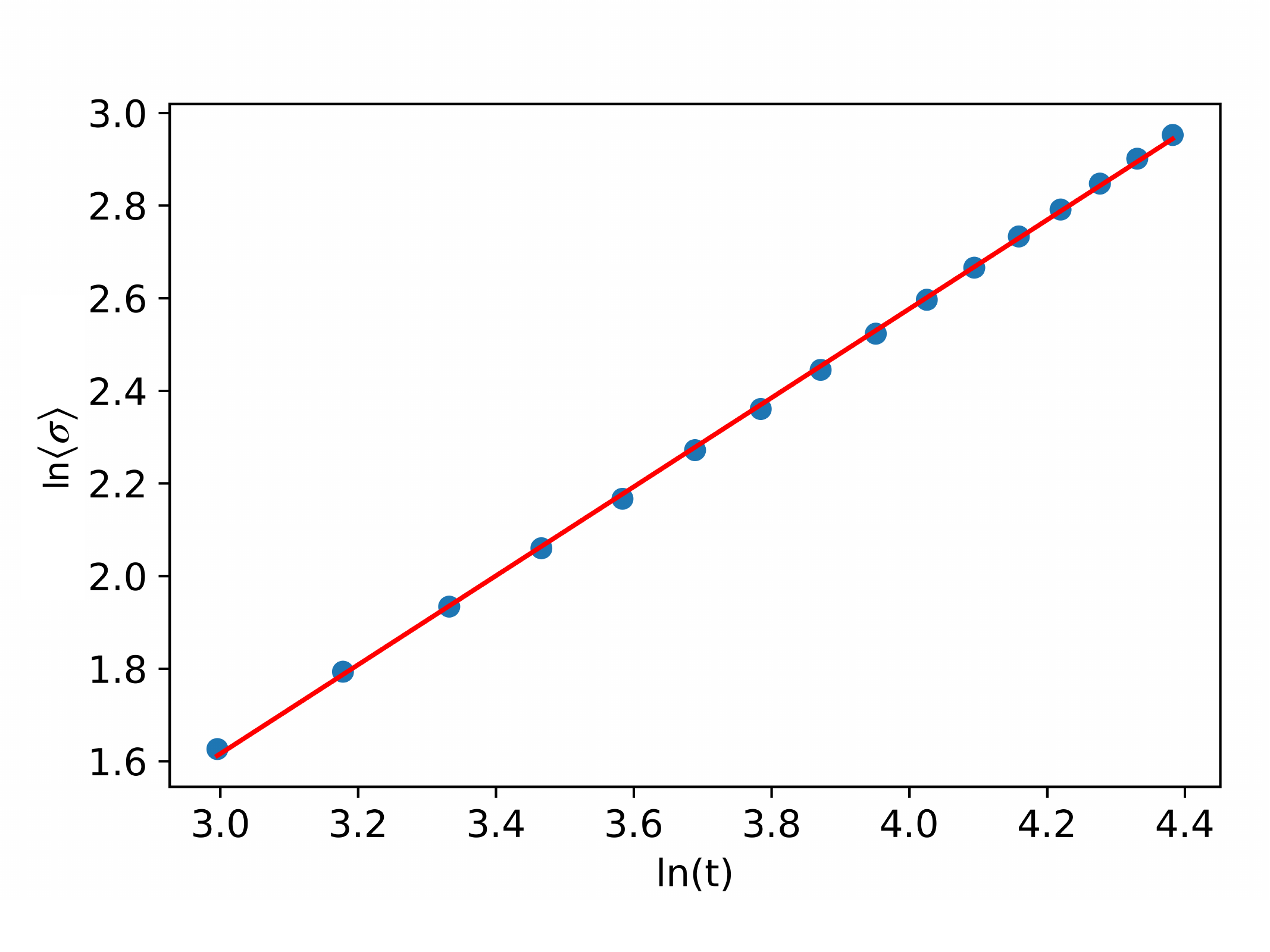}
    \caption{(Color online.) Plot of $\ln(\langle\sigma\rangle)$ vs $\ln(t)$ for the disordered-DTQW with one absorber placed at the position $2$ on the 1D line.
    The red solid line represents the linear fitting (on the log-log plot) for the data points (blue colored dots) corresponding to the Poisson disordered (mean $1$) 1D-DTQW.
    The slope of the fitted line lies within the confidence intervals $0.98\pm 0.004$, with $95\%$ confidence level,  with an average least square error of $0.002$ for the linear fittings. Here, $t$ varies from \(20\)  to \(80\). 
    }
    \label{fig: with_abs_with_dis}
\end{figure}

In the presence of an observer, the disordered CRWs for different probability distributions with different strengths show no significant changes in spreading rate compared to the clean walk or the walk in the absence of an absorber. For Poisson disordered CRWs with mean $1,$ a pictorial view in support of the last line is provided in Fig. \ref{fig: classical spread}. It is worth observing that both lines have an equal slope ($0.51$), which signifies the diffusive nature of the corresponding walks as like a clean CRW with no absorber case. In contrast, when an absorber is added to a disordered DTQW regime, the spreading rate is enhanced, and the walk almost restores the ballistic nature of a clean DTQW. First, we start analyzing this phenomenon with the Poisson distributed disorders of unit mean. Fig. \ref{fig: with_abs_with_dis} depicts the change of $\ln(\langle\sigma\rangle)$ along the vertical direction against $\ln(t)$ along the horizontal. 
Here, to calculate $\langle \sigma \rangle$ we take the average over $1100$ disorder realization sets.
In this plot, we use a log-log (natural log) scale for better understanding of the relation. Using the method of least squares, we fit a straight line to the data set. The slope of the straight line is evaluated as $0.98\pm0.004,$ which is the value of the exponent $\alpha.$  Here, the numerical figures are corrected up to two significant figures; for more details on the process, we refer to \cite{mandal2023stronger}. We continue the process for other Poisson disorders with different strengths. This signifies that after the insertion of an absorber along the walker's line, a Poisson disordered DTQW, which was sub-ballistic and super-diffusive initially, regains the ballistic-like spreading as that of a clean DTQW.



\begin{table}[]
\centering
\begin{tabular}{|c|c|c|c|}
\hline Distribution & Variance & Absorber present & $\alpha$\\
\hline
\hline
\multirow{3}{*} {Binomial} & $\frac{1}{2}$ & yes & $1.02$ \\
\cline{3-4}
  & & no & $0.79$ \\
\cline{1-4} Hypergeometric & $\frac{4}{9}$ & yes & $1.03$ \\
\cline{3-4}
  & & no & $0.81$ \\
\hline
{Negative} & $2$ & yes & $0.91$ \\
\cline{3-4}
 {Binomial} & & no & $0.69$ \\
\cline{1-4}
{Geometric} & $2$ & yes & $1.01$ \\
\cline{3-4}
  & & no & $0.74$ \\
\hline
\end{tabular}
\caption{Numerically evaluated values of the exponent $\alpha$, corrected up to the second decimal places, for various discrete disorders with the distribution mean value $1,$ for all the cases. Corresponding variances for the distributions are listed in the second column. In all such instances, except with the negative binomial distribution,  $\alpha$ increases to $1$ approximately, after the insertion of an absorber on the walker's line of movement. For the negative binomial distribution, restoration of the $\alpha$ values towards the transition from sub-ballistic to ballistic, also shows a consistency with the other cases.
Thus, the sub-ballistic spreading of disordered DTQWs, with random step lengths following sub- or super-Poissonian distributions, turns into a ballistic-like nature. 
}
\label{table: Sub- and Super- possinian}
\end{table}

In TABLE \ref{table: Sub- and Super- possinian}, we provide a list of numerically evaluated exponent values of disordered DTQWs, for a variety of discrete probability distributions, all with unit mean value. We see that even if the distribution is different from the Poisson distribution, the presence of an absorber renders the effects of disorders irrelevant by restoring the rate of spreading of a disordered DTQW to that of a clean one.
\vspace{-0.5cm}
\section{conclusion} \label{sec:conclusion}

In this article, we focus on how the presence of an absorber affects a one-dimensional discrete-time quantum walk. First, we define the average absorbing time and analytically show that this quantity diverges for a 1D classical random walk for any arbitrary choice of the absorber's position. Then, we prove that the average absorbing time for a 1D discrete-time quantum walk converges when the absorber is placed at the node $2$. Also, for other positions of the absorber, we calculate the finite average absorbing time numerically. 
Moreover, if we induce disorders in the step lengths of the walks, this behavior of the walks is altered, i.e., the average absorbing time for CRW and DTQW converges and diverges, respectively. For a numerical establishment of the claim, we undertake the disordered step lengths that follow either of the Poisson, sub-Poisson, and super-Poisson discrete probability distributions.

Next, we discuss the influence of an absorber in the spreading and analyze the corresponding statistical properties of random walks. We observe that for the Poisson distributed disorders, the presence of an absorber enhances the spreading rate of the disordered quantum walker, whereas it remains unchanged in the classical regime.   
Furthermore, we generalize this concept to sub- and super-Poissonian disorders as well, with different strengths. In fact, in all such instances, due to the presence of an absorber, the sub-ballistic nature of disordered DTQWs almost turns into a ballistic one. Thus, the absorber renders step-length disorders insignificant by restoring the spreading as a clean walk. Studying the effects of the absorber on average absorbing time, spreading rate, etc., of DTQWs for disorders in the coin
operations, and whether the behavior persists as discussed in this paper, is a problem of future interest.


\begin{acknowledgments}
We acknowledge discussions with Subhasis Jena. We thank the cluster facilities at the Harish-Chandra Research Institute.  We acknowledge partial support from the Department of Science and Technology, Government of India through the QuEST grant (grant number DST/ICPS/QUST/Theme-3/2019/120).
\end{acknowledgments}
\vspace{1.5cm}

\bibliography{qw_refs}
\onecolumngrid
\appendix
\section{Glassy Disorders in DTQWs}\label{sec:glassy_disorder}
Before introducing the disorder in the walker's step length, the evolution of the quantum system was not considered random, and the randomness in the system was only in the results of measurement after a walk. Now, we introduce disorder in the evolution of the quantum system by making the length of the steps of the quantum walk random.
In an ordered system, the walk evolves by taking the same step lengths at all the time steps. However, in a disordered system, the insertion of disorder is carried out in the walker's steps, such that after each time step, the walker moves following a step of any random length ($l$) related to the probability function $\mbox{pmf}(r;l)$ under consideration.
In detail, a probability distribution of a variable outcome $l$ is outlined as $\mbox{pmf}(r;l)$, where $r$ denotes some suitable parameter of the given distribution.
Now the walker can take a step of length $l$ after each toss, unlike to the ordered situation, where the length of the step is taken to be $1$ after each toss. From a mathematical perspective, when we include disorder, the shift operator, which has to be operated after the operation of the coin tossing operator, takes the form 
 \begin{equation}\label{eq: def_disordered_shift_opr}
     \hat{S}=\sum_{n=-\infty}^{\infty}\ket{n-l,L}\bra{n,L}+\ket{n+l,R}\bra{n,R},
 \end{equation}
 unlike to Eq.~(\ref{eq: def_shift_opr}).
 Here, $l$ has to be chosen with specified probability $\mbox{pmf}(r;l)$ for some fixed parameters $r$.
 

In this paper, unless explicitly mentioned, we refer to the disorders employed in the system as \textit{glassy disorders}. As an essential feature of the glassy disorder, the parameter $r$ of the specified distribution $\mbox{pmf}(r;l)$ remains unchanged during a full walk. Below, we state the probability mass functions of the  Poissonian, sub- and super-Poissonian distributions whose discrete outcomes serve the role of disorder.

\textbf{Poisson distribution.}
In this distribution, the only parameter $\lambda$ represents the mean value of the distribution and the mass function $\mbox{pmf}(\lambda;l)$ is defined as
\beano
\mbox{pmf}(\lambda, l)=\frac{\lambda ^{l}e^{-\lambda}}{l!}.
\eeano
Both the mean value and the variance of such a distribution are equal to $\lambda$.


\textbf{Sub- and super-Poisson distributions.} \label{subsec: sub & super poissonian disorder}
The probability distribution for which the variance is larger (smaller) than a Poisson distribution with the same mean value is called a super- (sub-) Poissonian distribution. For example, binomial and hypergeometric are sub-Poissonian distributions, whereas  
 negative-binomial and geometric are super-Poissonian distributions.
 
  \textbf{Binomial distribution.} 
  The probability function for this distribution is 
 \beano
 \mbox{pmf}(n,p;l) = \binom{n}{l} p^l (1-p)^{n-l},
\eeano
where $n$ and $p$ are two free parameters of the distribution. The mean value and the variance of this distribution are $np$ and $np(1-p),$ respectively. Clearly, $np>np(1-p),$ since $0\leq p \leq 1$ is the probability value.

 \textbf{Hypergeometric distribution.} The probability mass function of the hypergeometric distribution is 
 \beano
 \mbox{pmf}(N,K,n;l) = \frac{\binom{K}{l} \binom{N-K}{n-l}}{\binom{N}{n}},
\eeano
with $N,K$, and $n$ as the free parameters. The mean value and the variance for the distribution are $\frac{nK}{N}$ and  $\frac{nK(N-K)(N-n)}{N^{2}(N-1)},$ respectively. 

  \textbf{Negative binomial distribution.} 
   The mass function of the specified distribution is 
 \beano
 \mbox{pmf}(r,k;l) = \binom{l+ r - 1}{l} (1 - k)^r k^l,
\eeano
with $r$ and $k$ as parameters of the distribution. The mean and variance are $\frac{r(1-k)}{k},$ and $\frac{r(1-k)}{k^{2}},$ respectively. 
 
\textbf{Geometric distribution.} 
 The probability mass function of the geometric distribution is 
 \beano
 \mbox{pmf}(k;l)= (1-k)^{l-1}k,
\eeano
where $k$ acts as a parameter. The mean and variance of this distribution can be derived as $\frac{1}{k}$ and $\frac{1-k}{k^{2}},$ respectively. 

\section{}\label{apndx B}
Here, we sketch the proof that the total absorption probability of a classical walker is $1$ and does not depend on the distance of the absorber from the starting point of the walker.

Let us assume that the absorber is at the position $-m_{1}$ (without any loss of generality, we can assume that $m_{1}\in \mathbb{Z}$ is positive). Let $P(-m_{1})$ be the total probability of absorption of the walker starting from zero, and $p_t(-m_{1})$ stands for the probability of absorption at  $-m_{1}$ for the first time after $t$ time steps. 
Clearly,
$P(-1)=\frac{1}{2}+\frac{1}{2}P(-1)P(-1),$ which yields $P(-1)=1$ \cite{kempe2003quantum}.
Now to calculate $P(-2)$, which looks as
\beano \frac{1}{2}\times \frac{1}{2}+\left(\frac{1}{2}\times \frac{1}{2}+\frac{1}{2}\times \frac{1}{2}\right) P(-2)+\frac{1}{2}\times \frac{1}{2} P(-2) P(-2),\eeano 
which simplifies to $P(-2)=1.$

Finally, consider $P(-m_{1}),$ which can be expressed as
\begin{eqnarray} \label{eq: total classical probability series}
   P(-m_{1}) =\sum_{i=0}^{\infty}p_{m_{1}+2i}(-m_{1}).
\end{eqnarray}
Now, the following recurrence relations can be obtained for $m_{1} > 2$, in a similar way as we calculated $P(-2)$.

\begin{align*}
    p_{m_{1}}(-m_{1})&=\frac{1}{4}p_{m_{1}-2}(-(m_{1}-2)),\\
p_{m_{1}+2}(-m_{1})&=\frac{1}{4} p_{m_{1}}(-(m_{1}-2)) + \frac{1}{2} p_{m_{1}}(-m_{1}),\\
p_{m_{1}+4}(-m_{1})&=\frac{1}{4} p_{m_{1}+2}(-(m_{1}-2)) + \frac{1}{2} p_{m_{1}+2}(-m_{1})  +\frac{1}{4}p_{2}(-2)p_{m_{1}}(m_{1}),\\
p_{m_{1}+6}(-m_{1})&=\frac{1}{4} p_{m_{1}+4}(-(m_{1}-2)) + \frac{1}{2} p_{m_{1}+4}(-m_{1}) +\frac{1}{4}\left(p_{4}(-2)p_{m_{1}}(m_{1})+p_{2}(-2)p_{m_{1}+2}(m_{1})\right).
\end{align*}

Using Eq.~(\ref{eq: total classical probability series}) and the above set of recurrence relations, we get 
\begin{equation}\label{eq: total classical absorption probability}
    P(-m_{1})=\frac{1}{4}P(-(m_{1}-2))+\frac{1}{2}P(-m_{1})+\frac{1}{4}P(-2)P(-m_{1}).
\end{equation}
From Eq.~(\ref{eq: total classical absorption probability}) and using $P(-2)=1,$ we get
\begin{equation}\label{eq: total classicaal probability recurrance}
    P(-m_{1})=P(-(m_{1}-2)).
\end{equation}
Hence, $P(-1)=1, P(-2)=1,$ and Eq.~(\ref{eq: total classicaal probability recurrance}) together imply that 
$P(-m_{1})=1$ for any $m_1>0.$
Note that, following a similar approach presented above, we can show that $P(-m_{1})=1$ whenever $m_{1}$ is a negative integer. 

\section{}\label{apndx C}
In this section, we prove the divergence of the series given in Eq.~(\ref{eq: average_absorbing_time_in_classical}) and the convergence of the series in Eq.~(\ref{eq: quantum_avg_time}). Before that, we briefly describe Raabe's Test for testing the convergence of a series of positive terms ~\cite{Arfken1985}. \\
\textbf{Raabe's Test}: Consider an infinite series $\sum_{n} u_{n}$, where  $u_n> 0$ for every $n.$ Then, the series $\sum_{n} u_{n},$ converges if $E > 1$, diverges if $E < 1,$ and no conclusion can be made if $E = 1,$
 where 
\begin{equation}\label{eq: raabe estimator}
    E=\lim_{n \to \infty}n\left(\frac{u_{n}}{u_{n+1}}-1\right).
\end{equation}


For convenience, we assume that $m_{1}$ is positive, then the series sum in Eq.~(\ref{eq: average_absorbing_time_in_classical}) can be written in the following form since $t\geq m_1.$ 
\[\sum_{n=1}^{\infty}\frac{m_{1}(m_{1}+2n)!}{2^{m_{1}+2n}(\frac{m_{1}+2n+m_{1}}{2})!(\frac{m_{1}+2n-m_{1}}{2})!}=\frac{m_{1}}{2^{m_{1}}}\sum_{n=1}^{\infty}\frac{(m_{1}+2n)!}{4^{n}(m_{1}+n)!\,n!}.\]
Now, let us calculate $E$ as described in Eq.~(\ref{eq: raabe estimator}) for the above expression.
\beano E&=&\lim_{n \to \infty}n\left(4\frac{(m_{1}+2n)!}{(m_{1}+2n+2)!}\frac{(n+1)!}{n!}\frac{(m_{1}+n+1)!}{(m_{1}+n)!}-1 \right)\\
&=&\lim_{n \to \infty}n\left(\frac{(n+1)(n+1+m_{1})}{(n+\frac{1+m_{1}}{2})(n+\frac{2+m_{1}}{2})}-1 \right)\\
&=&\lim_{n \to \infty}\frac{\frac{1}{2}+\frac{1+m_{1}}{n}-\frac{(1+m_{1})(2+m_{1})}{4n}}{1+\frac{3+2m_{1}}{2n}+\frac{(1+m_{1})(2+m_{1})}{4n^{2}}}\\
&=&\frac{1}{2} < 1.\eeano
Hence, the series mentioned in Eq.~(\ref{eq: average_absorbing_time_in_classical}) diverges by applying Raabe's test.

For the infinite series given in Eq.~(\ref{eq: quantum_avg_time}), we calculate $E$ as follows. 
\beano E&=&\lim_{n \to \infty}n\left( \frac{4n-2}{4n+2} \left( \frac{(2n-2)!}{(2n)!} \right)^{2} \left( \frac{n!(n+1)!}{(n-1)!n!} \right)^{2} \left( \frac{2^{2n+1}}{2^{2n-1}} \right)^{2}-1 \right)\\
\eeano
This can be simplified as
\begin{align*}
    E=\lim_{n \to \infty}n\left( 16 \left(\frac{4n-2}{4n+2}\right)\frac{n^{2}(n+1)^{2}}{(2n-1)^{2}(2n)^{2}}-1 \right)\\
\end{align*}
$$=\lim_{n \to \infty}n\left( \left(\frac{n-\frac{1}{2}}{n+\frac{1}{2}}\right)\frac{(n+1)^{2}}{(n-\frac{1}{2})^{2}}-1 \right)\\$$
$$E=\lim_{n \to \infty}n\left( \frac{(n+1)^{2}}{n^{2}-\frac{1}{4}} -1 \right)\\$$
$$=\lim_{n \to \infty}n\left( \frac{2n+\frac{5}{4}}{n^{2}-\frac{1}{4}} \right)=\lim_{n \to \infty}\left( \frac{2+\frac{5}{4n}}{1-\frac{1}{4n^{2}}} \right)=2>1$$

Thus, by Raabe's test, the series sum in Eq.~(\ref{eq: quantum_avg_time}) converges.

\end{document}